\documentclass[english,conference]{IEEEtran}
\usepackage[T1]{fontenc}
\usepackage[cmex10]{amsmath}
\usepackage[latin9]{inputenc}
\usepackage{xcolor}
\usepackage{pdfcolmk}
\usepackage{units}

\usepackage{amssymb}
\usepackage{epsfig}
\usepackage{stackrel}
\usepackage{graphicx}
\usepackage{esint}
\usepackage{cite}
\PassOptionsToPackage{normalem}{ulem}
\usepackage{ulem}

\makeatletter

\pdfpageheight\paperheight
\pdfpagewidth\paperwidth

\providecolor{lyxadded}{rgb}{0,0,1}
\providecolor{lyxdeleted}{rgb}{1,0,0}
\usepackage[a4paper, left=1.75cm, right=1.75cm, top=2.3cm, bottom=2.7cm]{geometry}

\makeatother

\usepackage{babel}
\begin{document}

\title{A New Cell Association Scheme In Heterogeneous Networks}
\author{\IEEEauthorblockN{Bin Yang\IEEEauthorrefmark{1},
Guoqiang Mao\IEEEauthorrefmark{4},
Xiaohu Ge\IEEEauthorrefmark{1},
Tao Han\IEEEauthorrefmark{1}
}
\IEEEauthorblockA{\IEEEauthorrefmark{1} School of Electronic Information \& Communications,
Huazhong University of Science \& Technology, Wuhan, China
}
\IEEEauthorblockA{\IEEEauthorrefmark{4} School of Computing and Communication,
University of Technology Sydney, Australia\\
Corresponding Author: Xiaohu Ge, Email: xhge@mail.hust.edu.cn}
}
\maketitle
\begin{abstract}

Cell association scheme determines which base station (BS) and mobile
user (MU) should be associated with and also plays a significant role
in determining the average data rate a MU can achieve in heterogeneous networks. However, the explosion
of digital devices and the scarcity of spectra collectively force
us to carefully re-design cell association scheme which was kind of
taken for granted before. To address this, we develop a new cell association
scheme in heterogeneous networks based on joint consideration of the
signal-to-interference-plus-noise ratio (SINR) which a MU experiences
and the traffic load of candidate BSs%
\footnote{Candidate BSs is comprehended as the set of BSs which a mobile user (MU) is most
likely associated with.%
}. MUs and BSs in each tier are modeled
as several independent Poisson point processes (PPPs) and all channels
experience independently and identically distributed (\emph{i.i.d.}) Rayleigh fading. Data rate ratio and traffic load ratio distributions
are derived to obtain the tier association probability and the average
ergodic MU data rate. Through numerical results, We find that our proposed cell association scheme outperforms cell range expansion (CRE) association scheme. Moreover, results indicate that allocating small sized and high-density BSs will improve spectral efficiency if using our proposed cell association scheme in heterogeneous networks.

\end{abstract}
\vspace{2 ex}

\section{Introduction\label{sec:Introduction}}

Driven by a new revolution of digital devices like smart phones, tablets
and so on, there has been experiencing a tremendous growth of mobile
internet traffic in recent years. Traditional network expansion techniques
like cell splitting are often utilized by telecom operators to achieve
the expected throughput, which are less efficient and proven not to
keep up with the pace of traffic proliferation in the near future.
Heterogeneous networks then become a promising and attractive network
architecture to settle this. Heterogeneous networks are a broad term
that refers to the coexistence of different networks (e.g., traditional
macrocell and small-cell networks like femtocells and picocells),
each of them constituting a network tier. Due to differences in deployment,
base stations (BSs) in different tiers may have or use different transmission power levels,
radio access technologies, fading environments and spatial densities.
Heterogeneous networks are envisioned to cope with most problems of
existing network architectures like dead spots, inter-cell interference,
less efficient, etc and has been introduced in the LTE-Advanced standardization
\cite{Hu2013Heterogeneous}. Massive work has been done in heterogeneous networks scenario mainly related
with coverage modeling\cite{ElSawy14Two,Dhillon12Modeling}, cooperative communications \cite{Li12Cooperative}, energy
consumption modeling \cite{YongSheng13Energy,Hu2014An}, interference
cancellation \cite{Jemin13Spectrum}, interference management \cite{Chun-Hung12Ergodic} and resource allocation \cite{Elsherif13Resource,Qian13Intracell,Singh13Joint}, however none of which pays enough
attention to existing problems on cell association schemes.

\subsection{Motivation and related work\label{sec:Related-work}}

In heterogeneous cellular networks, there are more BSs which a MU
can choose to be associated with than in traditional homogeneous single-tier
cellular networks. Therefore, cell association scheme is an indispensable
factor in wireless networks modeling. By using the maximum received signal strength (RSS) as cell association scheme, ElSawy and Hossain
quantified the performance gain in the outage probability obtained by
introducing cognition into femtocells in two-tier heterogeneous
networks \cite{ElSawy14Two}. In \cite{Ali11Performance}, Ali and Saquib developed a
practical yet tractable method of evaluating vertical handover algorithms
in a WLAN/Cellular two-tier heterogeneous network and the cell association
is also based on the maximum RSS. Dhillon \textit{et al.} \cite{Dhillon12Modeling} proposed a tractable
and accurate model for a downlink heterogeneous cellular network consisting
of K tiers of randomly located BSs. Novlan
\textit{et al.} aimed to evaluate two fractional frequency reuse (FFR) methods -- strict FFR and soft frequency reuse by using Poisson point processes (PPPs) in \cite{Novlan11Analytical}.
The cell association scheme utilized by Dhillon and Novlan is based on the maximum
downlink  signal-to-interference-plus-noise ratio (SINR). Also, the nearest BS cell association scheme is applied
in some literatures like \cite{YongSheng13Energy,Mukherjee12Distribution}. Yong Sheng \textit{et al.} investigated the design
and the associated tradeoffs of energy efficient heterogeneous cellular
networks through the deployment of sleeping strategies in \cite{YongSheng13Energy}.
In \cite{Mukherjee12Distribution}, Mukherjee provided a general theoretical
analysis of the distribution of the SINR at an arbitrarily-located
MU in heterogeneous networks.

From above literatures, existing cell association
schemes have been mainly based on the RSS, SINR or the distance from nearby BSs to determine
which BS and MU should be connected with each other. This is legitimate for traditional
homogeneous single-tier cellular networks where the RSS or the SINR
serves as a good indicator of the data rate received by the MU. However,
it is no longer the case in heterogeneous networks in which BSs from
different tiers transmit wireless signals at very different power levels, varying from
milliWatt (mW) to Watt (W): \textbf{\emph{a)}} the higher RSS may be a result of the higher
transmission power used by the BS. It may cause congestions
in BSs which have higher RSS and idleness in BSs whose RSS is lower
whereas can still guarantee successful transmission. This result brings
unbalance and inequity among BSs in different tier networks; \textbf{\emph{b)}} the number
of MUs served by a small-cell BS is typically small due to its much
smaller coverage. Consequently, the current traffic load of the BS
plays a significant role in determining the share of BS capacity received
by each MU. For example, the joining of a MU into a small-cell BS
currently serving one MU may halve the data rate received by the current MU; \textbf{\emph{c)}} as for choosing the nearest BS for association, it is so impractical that only used for
theoretical analysis. Thus, it is no longer optimum to determine cell association
solely based on the RSS, the SINR or the distance from nearby BSs.

As described above, cell association schemes play an important role
in determining the allocation of spectral resource in BSs, the transmission
rate that a MU can achieve and even the energy consumption of MUs. In \cite{Han-Shin12Heterogeneous,Lima13Statistical},
a solution was proposed to partially solve the problem \textbf{\emph{a)}} by introducing
a biased factor $\Omega$ or $B$ into the RSS, which allows an expansion
of the coverage of small-cell BSs. The effectiveness of the scheme
however remains questionable in networks with inhomogeneous user density,
e.g. MUs clustering around BSs. In \cite{Singh13Offloading}, authors mentioned the problem \textbf{\emph{b)}} in the subsection of resource allocation.
However, per MU data rate is only a performance metric with the form of
rate coverage and the used cell association scheme was still conventional,
which left these problems unsolved.

\subsection{Contributions and organization\label{sub:Contributions-and-organization}}

To solve problems \textbf{\emph{a)}}, \textbf{\emph{b)}} and \textbf{\emph{c)}}, a spectrum efficient
cell association scheme based on the joint consideration of the received SINR and the traffic load of BSs is proposed for heterogeneous networks. To match real BSs deployment scenarios, PPP is used to model heterogeneous cellular networks in this article, which has been strengthened by the empirical
validation \cite{Andrews11ATractable} and the theoretical validation
\cite{Blaszczyszyn13Using}. The contributions and novelties of this paper are summarized as follows.
\begin{enumerate}
\item A new cell association scheme is proposed with two steps for heterogeneous networks. The first step is mainly for choosing
the candidate BSs by traditional method, i.e., the nearest $n$ BSs,
while the second step determines the ultimate one BS based on the
consideration of the received SINR experienced by a MU and the traffic load of
candidate BSs.
\item Following the cell association scheme
and taking a three-tier heterogeneous network as an example, the tier association probability and the average ergodic MU data rate are derived for numerical analysis.
\item Based on numerical results, the new cell association scheme outperforms CRE association scheme.
\end{enumerate}

The reminder of this paper is organized as follows. In section \ref{sec:Network-model-and},
we present our network model and propose a new cell association scheme
in general case. A three-tier heterogeneous network is analyzed in section \ref{sec:Analysis-of-special}. Moreover, the tier association probability of heterogeneous networks is derived for performance analysis.
Section \ref{sec:Numerical-evaluation} presents the numerical results
of the proposed cell association scheme. Section \ref{sec:Conclusion-and-future} concludes this paper.

\section{Network model and proposed cell association scheme\label{sec:Network-model-and}}

We Consider a K-tier heterogeneous downlink cellular network which
consists of macrocells, picocells, femtocells, etc. BSs of each tier are assumed to be spatially distributed following
independent homogenous PPPs denoted by $\Phi_{k}$, $k\in\left\{ 1,2,\cdots,K\right\} $.
The BS intensity of the $k$-th tier network is $\lambda_{k}$, $k\in\left\{ 1,2,\cdots,K\right\} $.
MUs are located according to a homogeneous point process denoted
by $\Phi_{u}$ with intensity $\lambda_{u}$. All BSs in the same tier network
are configured with the same transmission power $P_{k}$, $k\in\left\{ 1,2,\cdots,K\right\} $
and share the same bandwidth. BSs in different tier networks are configured with different bandwidths. Moreover, within a cell, MSs are allocated by orthogonal
frequencies. Therefore, there is no intra-cell interference in a cell.
Also for simplicity, the open access policy is applied for MUs. It means all MUs can be served by BSs in any tier networks.

We propose a new cell association scheme that bases its cell
association decision on the instant traffic load of each BS and the transmission rate that
can be allocated by the BS. More specifically, the cell association
scheme can be divided into two steps.
\begin{enumerate}
\item If a MU wants to be associated with a BS, it will firstly choose $n$
nearest BSs from each tier as the candidate BSs. The candidate
BS set is defined as $\Omega_{B}=\left\{ \left(k,i\right)|k\in\left\{ 1,2,\cdots,K\right\} ,i\in[1, n]\right\} $,
where $k$ is the $k$-th tier network in a heterogeneous network and $i$ is the $i$-th BS in the $n$
nearest BSs from the $k$-th tier network. For
example, $\left(1,3\right)$ represents the 3rd BS in the 1st tier network. The
total number of BSs in $\Omega_{B}$ is $nK$.
\item The MU will select a candidate BS from $\Omega_{B}$. This selected BS will send data to the MU with the maximum average transmission rate, i.e., $\frac{B_{k,i}}{N_{k,i}+1}\cdotp\ln\left(1+\textrm{SINR}_{k,i}\right)$, where $B_{k,i}$ and $N_{k,i}$ are the total bandwidth of a candidate BS in set $\Omega_{B}$ and the
instant BS traffic load, respectively. $\textrm{SINR}_{k,i}$ is the MU instant SINR associated with a BS $\left(k,i\right)$ in the set $\Omega_{B}$.
It is assumed that the total BS bandwidth is shared
equally among all associated MUs.%
\footnote{In our following analysis, the candidate BS $\left(k,i\right)$ can be denoted by $k$ when
$n=1$. However, we'll keep using $\left(k,i\right)$ for completeness and preciseness.%
}
\end{enumerate}

\section{Performance analysis\label{sec:Analysis-of-special}}

\begin{figure}
  \centering
  \includegraphics[width=6.9cm, height = 6.2cm, draft=false]{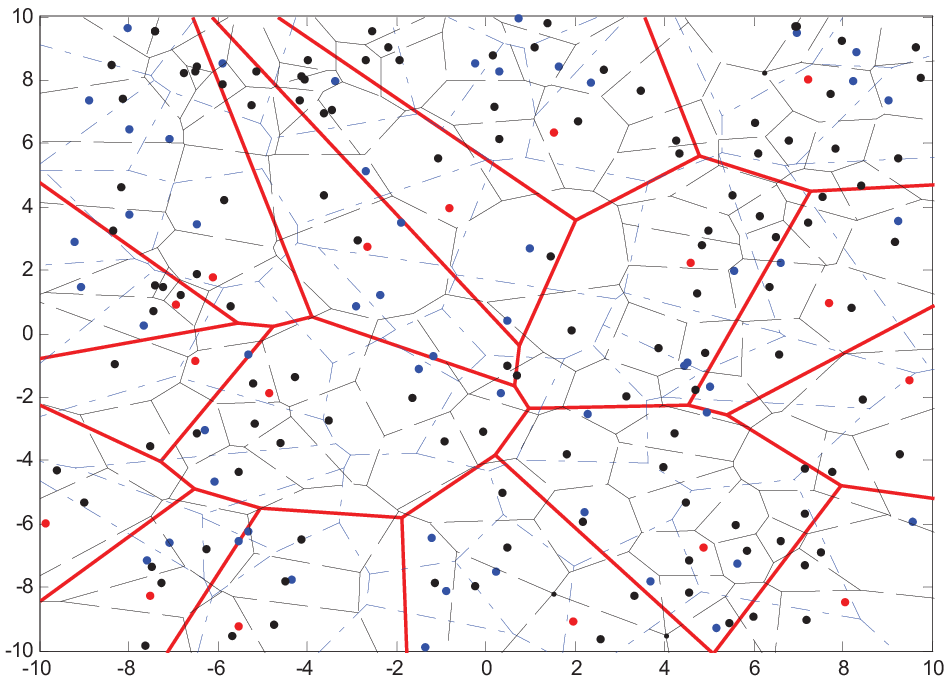}\\
  \caption{\small The three-tier heterogeneous network (20$km$$\times$20$km$
) modeled as a superposition of three independent Poisson Voronoi
tessellations. These polygons are 1st tier cells (edges with red solid lines),
2nd tier cells (edges with blue dot-dash lines) and 3rd tier cells
(edges with black dotted lines).}\label{fig1}
\end{figure}

Without generality, in the following we will analyse the scenario when $K=3$ and $n=1$. The network being considered
is a three-tier heterogeneous network and only the nearest BS at each
tier from a MU can be chosen as candidate BSs. The analysed three-tier heterogeneous
network is depicted in Fig. \ref{fig1}.

\subsection{The downlink SINR distribution\label{sub:The-downlink-SINR}}

For downlink transmission of a BS $\left(k,i\right)$ to a
MU which is located at the origin $o$, the SINR experienced by this MU is expressed by
\begin{equation}
\textrm{SINR}_{k,i}=\frac{P_{k}h_{k,i}r_{k,i}^{-\alpha}}{\underset{m\in\Omega_{k}^{'}}{\sum}P_{k}h_{k,m}r_{k,m}^{-\alpha}+\sigma^{2}},
\end{equation}
where $\Omega_{k}^{'}$ is set of interferers in the $k$-th tier network. $h_{k,i}$
and $h_{k,m}$ are channel power gains due to small-scale fading
between the considered MU and BS $\left(k,i\right)$, $\left(k,m\right)$, respectively. For convenience and without
generality, we assume $h_{k,i}\sim\exp\left(1\right)$ and $h_{k,m}\sim\exp\left(1\right)$.
The background noise is assumed to be additive white Gaussian with
variance $\sigma^{2}$. $r_{k,i}^{-\alpha}$ and $r_{k,m}^{-\alpha}$
are path losses with $\alpha$ being the path loss exponent, $r_{k,i}$
and $r_{k,m}$ being the respective Euclidean distance to the corresponding
BS $\left(k,i\right)$ and $\left(k,m\right)$, respectively.

Referring to \cite{YongSheng13Energy}, the coverage probability that
a MU is covered by its nearest BS in a particular tier $k$ is derived
as follows
\begin{align}
\boldsymbol{\textrm{P}_{c}^{k}}\left(x\right) & =\Pr\left(\textrm{SINR}_{k,i}>x\right)\nonumber \\
 & =2\pi\lambda_{k}\int_{r=0}^{+\infty}r\exp\left[-\pi r^{2}\lambda_{k}\left(1+\varphi\left(x\right)\right)\right]\times\label{eq:P_c}\\
 & \quad\exp\left(-\frac{r^{\alpha}x\sigma^{2}}{P_{k}}\right)dr,\nonumber
\end{align}
where $\varphi\left(x\right)=x^{\frac{2}{\alpha}}\int_{x^{-\frac{2}{\alpha}}}^{+\infty}\frac{1}{1+y^{\frac{\alpha}{2}}}dy$.
$\boldsymbol{\textrm{P}_{c}^{k}}\left(x\right)$ is the complementary cumulative distribution function (CCDF) of
$\textrm{SINR}_{k,i}$ and the cumulative distribution function (CDF) of $\textrm{SINR}_{k,i}$ is $1-\boldsymbol{\textrm{P}_{c}^{k}}\left(x\right)$.
By taking a derivative of the CDF of $\textrm{SINR}_{k,i}$ with respect to
$x$, the probability density function (PDF) of $\textrm{SINR}_{k,i}$
is obtained by

\begin{equation}
\begin{array}{cl}
f_{\textrm{SINR}_{k,i}}\left(x\right) & =2\pi\lambda_{k}\int_{r=0}^{+\infty}r\left[\pi r^{2}\lambda_{k}\varphi'\left(x\right)+\frac{r^{\alpha}\sigma^{2}}{P_{k}}\right]\times\\
 & \quad\exp\left[-\pi r^{2}\lambda_{k}\left(1+\varphi\left(x\right)\right)\right]\times\\
 & \quad\exp\left(-\frac{r^{\alpha}x\sigma^{2}}{P_{k}}\right)dr
\end{array}\label{eq:f_SINR}
\end{equation}
with

\begin{equation}
\varphi'\left(x\right)=\frac{2}{\alpha}\left[\frac{\varphi\left(x\right)}{x}+\frac{1}{1+x}\right].
\end{equation}

\subsection{The tier association probability\label{sub:Tier-association-probability}}

The spatial average data rate a MU can achieve is denoted by $\overline{C_{k}}$, $k\in\left\{ 1,2,3\right\} $
 and the event that the considered MU is associated with a BS in the
$k$-th tier network is denoted by $(N_{Tier}=k)$.
We'll apply Slivnyak\textquoteright s theorem to
the following analysis on a MU that located at the origin $o$, which
implies that conditioning on having that user at the origin, properties
of all coexisting PPPs maintaining the same. Using the above association
scheme in section \ref{sec:Network-model-and}, the probability that a MU is associated with the
1st tier BS $\left(1,i\right)$ is
\begin{equation}
\begin{array}{cl}
\mathcal{T}_{1} & =\Pr\left(N_{Tier}=1\right)\\
 & =\Pr\left(\overline{C_{1}}>\underset{j\neq1}{\max}\overline{C_{j}}\right)\\
 & =\Pr[\frac{B_{1,i}}{N_{1,i}+1}\ln\left(1+\textrm{SINR}_{1,i}\right)>\\
 & \quad\frac{B_{2,i}}{N_{2,i}+1}\ln\left(1+\textrm{SINR}_{2,i}\right),\frac{B_{1,i}}{N_{1,i}+1}\ln\left(1+\textrm{SINR}_{1,i}\right)\\
 & \quad >\frac{B_{3,i}}{N_{3,i}+1}\ln\left(1+\textrm{SINR}_{3,i}\right)]\\
 & \overset{\left(\text{\mbox{I}}\right)}{=}\stackrel[j=2]{3}{\prod}\Pr[\frac{B_{1,i}}{N_{1,i}+1}\ln\left(1+\textrm{SINR}_{1,i}\right)>\\
 & \quad\frac{B_{j,i}}{N_{j,i}+1}\ln\left(1+\textrm{SINR}_{j,i}\right)]\\
 & =\stackrel[j=2]{3}{\prod}\Pr\left[\frac{N_{1,i}+1}{N_{j,i}+1}<\frac{B_{1,i}}{B_{j,i}}\cdotp\frac{\ln\left(1+\textrm{SINR}_{1,i}\right)}{\ln\left(1+\textrm{SINR}_{j,i}\right)}\right]\\
 & \overset{\left(\text{\mbox{II}}\right)}{=}\stackrel[j=2]{3}{\prod}\Pr\left[N_{1/j}<\frac{B_{1,i}}{B_{j,i}}\cdotp\textrm{SINR}_{1/j}\right]\\
 & \overset{\left(\text{\mbox{III}}\right)}{=}\stackrel[j=2]{3}{\prod}\int_{0}^{+\infty}F_{N_{1/j}}\left(\frac{B_{1,i}}{B_{j,i}}\cdotp x\right)\cdotp f_{\textrm{SINR}_{1/j}}\left(x\right)dx
\end{array},\label{eq:T_1}
\end{equation}
where (\mbox{I}) is due to the independence between two events
$\left\{ \overline{C_{1}}>\overline{C_{2}}\right\} $ and $\left\{ \overline{C_{1}}>\overline{C_{3}}\right\} $;
in (\mbox{II}), $N_{1/j}$, $\textrm{SINR}_{1/j}$ denote $\frac{N_{1,i}+1}{N_{j,i}+1}$
and $\frac{\ln\left(1+\textrm{SINR}_{1,i}\right)}{\ln\left(1+\textrm{SINR}_{j,i}\right)}$,
respectively; (\mbox{III}) is obtained by applying the law of total
probability where $F_{N_{1/j}}\left(x\right)$ is the CDF of $\frac{N_{1,i}+1}{N_{j,i}+1}$ and
$f_{\textrm{SINR}_{1/j}}\left(x\right)$ denotes the PDF of $\frac{\ln\left(1+\textrm{SINR}_{1,i}\right)}{\ln\left(1+\textrm{SINR}_{j,i}\right)}$.

\subsubsection{The CDF of $\frac{N_{1,i}+1}{N_{j,i}+1}$}

In this paper, it is assumed that each BS has a unique saturated downlink
transmission queue for each MU. This assumption implies that MU always
has data to receive from a BS which covers and associates that MU\footnote{The shape of a BS's coverage is Voronoi-tessellated.%
}. Each MU choose the
associating BS with probabilities denoted by $\mathcal{T}_{k}$, $k\in\left\{ 1,2,3\right\} $. Thus, three point
processes denoted by $\Phi_{u}^{1}$, $\Phi_{u}^{2}$ and $\Phi_{u}^{3}$
are formed by thinning the original PPP $\Phi_{u}$. The thinned processes
are the locations of MUs which are associated with the 1st, the
2nd and the 3rd tier BSs. The thinned point processes are still
PPPs and intensities are $\mathcal{T}_{1}\lambda_{u}$, $\mathcal{T}_{2}\lambda_{u}$ and $\mathcal{T}_{3}\lambda_{u}$, respectively.

Through interpretations above, the probability mass function (PMF)
of $N_{k,i}$%
\footnote{ in this paper, traffic load of a BS is defined as the total number of MUs associated with that BS.%
} is given by
\begin{equation}
\begin{array}{cl}
f_{N_{k,i}}\left(n\right) & =\Pr\left(N_{k,i}=n\right)\\
 & =\int_{0}^{+\infty}\frac{\left(\mathcal{T}_{k}\lambda_{u}s\right)^{n}e^{-\mathcal{T}_{k}\lambda_{u}s}}{n!}\cdotp f_{S_{k}}\left(s\right)ds\\
\\
 & =\int_{0}^{+\infty}\frac{\left(\mathcal{T}_{k}\lambda_{u}s\right)^{n}e^{-\mathcal{T}_{k}\lambda_{u}s}}{n!}\cdotp\frac{\left(c\lambda_{k}\right)^{c}s^{c-1}e^{-c\lambda_{k}s}}{\Gamma\left(c\right)}ds\\
\\
 & =\frac{\left(\mathcal{T}_{k}\lambda_{u}\right)^{n}}{n!}\cdotp\frac{\left(c\lambda_{k}\right)^{c}}{\Gamma\left(c\right)}\times\\
\\
 & \int_{0}^{+\infty}\frac{\left(\mathcal{T}_{k}\lambda_{u}s\right)^{n}e^{-\mathcal{T}_{k}\lambda_{u}s}}{n!}s^{n+c-1}e^{-s\left(\mathcal{T}_{k}\lambda_{u}+c\lambda_{k}\right)}ds\\
\\
 & =\frac{\left(\mathcal{T}_{k}\lambda_{u}\right)^{n}\left(c\lambda_{k}\right)^{c}}{n!\Gamma\left(c\right)}\cdotp\frac{\Gamma\left(n+c\right)}{\left(\mathcal{T}_{k}\lambda_{u}+c\lambda_{k}\right)^{n+c}}\times\\
\\
 & \int_{0}^{+\infty}\underline{\frac{\left(\mathcal{T}_{k}\lambda_{u}+c\lambda_{k}\right)^{n+c}}{\Gamma\left(n+c\right)}\cdotp s^{n+c-1}\cdotp e^{-s\left(\mathcal{T}_{k}\lambda_{u}+c\lambda_{k}\right)}ds}\\
\\
 & \overset{\left(\text{\mbox{I}}\right)}{=}\frac{\left(\mathcal{T}_{k}\lambda_{u}\right)^{n}\left(c\lambda_{k}\right)^{c}}{\left(\mathcal{T}_{k}\lambda_{u}+c\lambda_{k}\right)^{n+c}}\cdotp\frac{\Gamma\left(n+c\right)}{\Gamma\left(n+1\right)\Gamma\left(c\right)}
\end{array},
\end{equation}
where $\Gamma\left(\cdotp\right)$ is the Gamma function; $f_{S_{k}}\left(s\right)\approx\frac{\left(c\lambda_{k}\right)^{c}s^{c-1}e^{-c\lambda_{k}s}}{\Gamma\left(c\right)}$
is the PDF of the Voronoi cell area of the $k$-th tier network obtained through
simulations and $c=3.575$ is a constant \cite{Ferenc07On}. (\mbox{I})
is obtained due to the integration of the formula with underline is 1 over the domain.
Actually, the format of the formula with underline is the PDF of Gamma distribution like $y=\frac{x^{k-1}e^{-\nicefrac{x}{\theta}}}{\theta^{k}\Gamma\left(k\right)}$.

The CDF of $\frac{N_{1,i}+1}{N_{j,i}+1}$ is derived by
\begin{equation}
\begin{array}{cl}
F_{N_{1/j}}\left(x\right) & =\Pr\left(\frac{N_{1,i}+1}{N_{j,i}+1}<x\right)\\
 & =\Pr\left[N_{1,i}<\left(N_{j,i}+1\right)x-1\right]\\
 & \overset{\left(\text{\mbox{I}}\right)}{=}\stackrel[t=0]{\infty}{\sum}\Pr\left[N_{1,i}<\left(t+1\right)x-1\mid N_{j,i}=t\right]\times\\
 & \quad\Pr\left(N_{j,i}=t\right)\\
 & =\stackrel[t=0]{\infty}{\sum}F_{N_{1,i}}\left[\lfloor\left(t+1\right)x-1\rfloor\right]\cdotp f_{N_{j,i}}\left(t\right)
\end{array},\label{eq:F_N/}
\end{equation}
where $F_{N_{k,i}}\left(l\right)$ is the CDF of $N_{k,i}$ which is derived
by
\begin{equation}
\begin{array}{cl}
F_{N_{k,i}}\left(l\right) & =\stackrel[n=0]{l}{\sum}f_{N_{k,i}}\left(n\right)\\
 & =\stackrel[n=0]{l}{\sum}\frac{\left(\mathcal{T}_{k}\lambda_{u}\right)^{n}\left(c\lambda_{k}\right)^{c}}{\left(\mathcal{T}_{k}\lambda_{u}+c\lambda_{k}\right)^{n+c}}\cdotp\frac{\Gamma\left(n+c\right)}{\Gamma\left(n+1\right)\Gamma\left(c\right)},\, l\in[0,\infty)
\end{array},\label{eq:F_N}
\end{equation}
$\lfloor\cdotp\rfloor$ is the floor function and (\mbox{I}) follows
the law of total probability. Substituting (\ref{eq:F_N}) back in
(\ref{eq:F_N/}), we obtain the CDF of $\frac{N_{1,i}+1}{N_{j,i}+1}$ as follows
\begin{equation}
\begin{array}{cl}
F_{N_{1/j}}\left(x\right) & =\stackrel[t=0]{\infty}{\sum}\{\frac{\left(\mathcal{T}_{j}\lambda_{u}\right)^{t}\left(c\lambda_{j}\right)^{c}}{\left(\mathcal{T}_{j}\lambda_{u}+c\lambda_{j}\right)^{t+c}}\cdotp\frac{\Gamma\left(t+c\right)}{\Gamma\left(t+1\right)\Gamma\left(c\right)}\times\\
 & \stackrel[n=0]{\lfloor\left(t+1\right)x-1\rfloor}{\sum}\frac{\left(\mathcal{T}_{1}\lambda_{u}\right)^{n}\left(c\lambda_{1}\right)^{c}}{\left(\mathcal{T}_{1}\lambda_{u}+c\lambda_{1}\right)^{n+c}}\cdotp\frac{\Gamma\left(n+c\right)}{\Gamma\left(n+1\right)\Gamma\left(c\right)}\}
\end{array}.
\end{equation}

\subsubsection{The PDF of $\frac{\ln\left(1+\textrm{SINR}_{1,i}\right)}{\ln\left(1+\textrm{SINR}_{j,i}\right)}$}

Let the PDF of $\textrm{SINR}_{k,i}$ be $f_{\textrm{SINR}_{k,i}}\left(x\right)$
and the CDF be $F_{\textrm{SINR}_{k,i}}\left(x\right)$, then the
CDF of $\ln\left(1+\textrm{SINR}_{k,i}\right)$ is derived by
\begin{equation}
\begin{array}{cl}
F_{\ln\left(1+\textrm{SINR}_{k,i}\right)}\left(y\right) & =\Pr\left[\ln\left(1+\textrm{SINR}_{k,i}\right)<y\right]\\
 & =\Pr\left(\textrm{SINR}_{k,i}<e^{y}-1\right)\\
 & =F_{\textrm{SINR}_{k,i}}\left(e^{y}-1\right)
\end{array}.\label{eq:F_ln}
\end{equation}
By taking a derivative with respect to $y$ in both sides of (\ref{eq:F_ln}), the PDF of $\ln\left(1+\textrm{SINR}_{k,i}\right)$ is obtained by
\begin{equation}
f_{\ln\left(1+\textrm{SINR}_{k,i}\right)}\left(y\right)=e^{y}\cdotp f_{\textrm{SINR}_{k,i}}\left(e^{y}-1\right).
\end{equation}
Let $f_{jo}\left(x,y\right)$ denote the joint probability density
function (JPDF) of random variable tuple $\left(\ln\left(1+\textrm{SINR}_{1,i}\right),\ln\left(1+\textrm{SINR}_{j,i}\right)\right)$,
$j\in\left\{ 2,3\right\} $. The PDF of $\frac{\ln\left(1+\textrm{SINR}_{1,i}\right)}{\ln\left(1+\textrm{SINR}_{j,i}\right)}$
is derived by
\begin{equation}
\begin{array}{cl}
f_{\textrm{SINR}_{1/j}}\left(z\right) & =\int_{-\infty}^{\infty}|y|\cdotp f_{jo}\left(zy,y\right)dy\\
 & \overset{\left(\text{\mbox{I}}\right)}{=}\int_{0}^{\infty}y\cdotp f_{jo}\left(zy,y\right)dy
\end{array},\label{eq:f_SINR/}
\end{equation}
where (\mbox{I}) is obtained by using the ratio distribution (or quotient distribution)
formula of two nonegative random variables. Because of the independence
of the two variables, i.e., $\ln\left(1+\textrm{SINR}_{1,i}\right)$
and $\ln\left(1+\textrm{SINR}_{j,i}\right)$,
$f_{jo}\left(x,y\right)$ is expressed by
\begin{equation}
\begin{array}{cl}
f_{jo}\left(x,y\right) & =f_{\ln\left(1+\textrm{SINR}_{1,i}\right)}\left(x\right)\cdotp f_{\ln\left(1+\textrm{SINR}_{j,i}\right)}\left(y\right)\\
 & =e^{x+y}f_{\textrm{SINR}_{1,i}}\left(e^{x}-1\right)\cdotp f_{\textrm{SINR}_{j,i}}\left(e^{y}-1\right)
\end{array}.\label{eq:f_jo}
\end{equation}
Substituting (\ref{eq:f_jo}) back into (\ref{eq:f_SINR/}), we obtain the PDF of $\frac{\ln\left(1+\textrm{SINR}_{1,i}\right)}{\ln\left(1+\textrm{SINR}_{j,i}\right)}$ as follows
\begin{equation}
\begin{array}{cl}
f_{\textrm{SINR}_{1/j}}\left(z\right) & =\int_{0}^{\infty}ye^{\left(z+1\right)y}f_{\textrm{SINR}_{1,i}}\left(e^{zy}-1\right)\times\\
 & \quad f_{\textrm{SINR}_{j,i}}\left(e^{y}-1\right)dy
\end{array},
\end{equation}
where $f_{\textrm{SINR}_{k,i}}\left(x\right)$ is given by (\ref{eq:f_SINR}).

Similarly, by repeating the derivations above, the probabilities
that a MU is associated with a BS in the 2nd and the 3rd tier network are obtained, i.e.,
\begin{equation}
\begin{array}{cl}
\mathcal{T}_{2} & =\stackrel[j=1,j\neq2]{3}{\prod}\int_{0}^{+\infty}F_{N_{2/j}}\left(\frac{B_{2,i}}{B_{j,i}}\cdotp x\right)\cdotp f_{\textrm{SINR}_{2/j}}\left(x\right)dx\end{array},\label{eq:T_2}
\end{equation}
\begin{equation}
\begin{array}{cl}
\mathcal{T}_{3} & =\stackrel[j=1]{2}{\prod}\int_{0}^{+\infty}F_{N_{3/j}}\left(\frac{B_{3,i}}{B_{j,i}}\cdotp x\right)\cdotp f_{\textrm{SINR}_{3/j}}\left(x\right)dx\end{array}.\label{eq:T_3}
\end{equation}
Using numerical method, the exact value of $\mathcal{T}_{k}$, $k\in\left\{ 1,2,3\right\} $ are obtained
by solving equations (\ref{eq:T_1}), (\ref{eq:T_2}) and (\ref{eq:T_3}).

\subsection{The average ergodic MU data rate\label{sub:The-average-ergodic}}

In this subsection, we focus on the average ergodic MU data rate of a MU
in a K-tier heterogeneous network. We assume that Shannon's capacity
can be achieved by some coding methods. The average ergodic MU data rate can be
obtained by considering per tier user data rates weighted by the corresponding
tier association probabilities. The average ergodic MU rate in a 3-tier heterogeneous
network is given by
\begin{equation}
\overline{\Re}=\stackrel[k=1]{3}{\sum}\mathcal{T}_{k}\Re_{k}.\label{eq:R}
\end{equation}
$\Re_{k}$ is the ergodic MU data rate conditioning on a MU is associated
with a specific BS in the $k$-th tier network which is given by
\begin{equation}
\begin{array}{rl}
\Re_{k} & =\mathbf{\boldsymbol{\mathbf{E}}}\left[B_{k,i}\cdotp\ln\left(1+\textrm{SINR}_{k,i}\right)\right]\\
 & \overset{\left(\text{\mbox{I}}\right)}{=}B_{k,i}\int_{0}^{\infty}\Pr\left[\ln\left(1+\textrm{SINR}_{k,i}\right)>t\right]dt\\
 & =B_{k,i}\int_{0}^{\infty}\boldsymbol{\textrm{P}_{c}^{k}}\left(e^{t}-1\right)dt
\end{array},\label{eq:R_k}
\end{equation}
where (\mbox{I}) is derived because $\ln\left(1+\textrm{SINR}_{k,i}\right)$
is a nonnegative random variable; $\mathbf{\boldsymbol{\mathbf{E}}\left(\cdotp\right)}$
is an expectation operator and $\boldsymbol{\textrm{P}_{c}^{k}}\left(\cdotp\right)$ is given by (\ref{eq:P_c}). Substituting (\ref{eq:R_k}) into (\ref{eq:R}),
we can get the unconditional average ergodic MU data rate as follows
\begin{equation}
\begin{array}{rl}
\overline{\Re} & =\stackrel[k=1]{3}{\sum}2\pi\lambda_{k}\mathcal{T}_{k}B_{k,i}\int_{0}^{\infty}\int_{r=0}^{+\infty}r\times\\
 & \quad\exp\left[-\pi r^{2}\lambda_{k}\left(1+\varphi\left(e^{t}-1\right)\right)\right]\times\\
 & \quad\exp\left(-\frac{r^{\alpha}\sigma^{2}\left(e^{t}-1\right)}{P_{k}}\right)drdt
\end{array}.
\end{equation}

\section{Numerical results and discussions\label{sec:Numerical-evaluation}}

This section presents numerical results of previous sections, followed
by discussions. Parameters used in this article are
refereed to existing work focused on heterogeneous networks. Specifically, we assume that $\sigma^{2}=0$ which denotes a interference-limited scenario. BS densities and BS transmission powers are $\lambda_{2}=2\lambda_{1}$,
$\lambda_{3}=20\lambda_{1}$, $\lambda_{u}=50\lambda_{1}$, $P_{1}=53\textrm{dBm}$,
$P_{2}=33\textrm{dBm}$, $P_{3}=23\textrm{dBm}$ \cite{Han-Shin12Heterogeneous,Singh13Offloading}. Allocations of spectra are divided into 4 cases, i.e., \textbf{(B1>B2>B3):} $B_{1}=15\textrm{MHz}$, $B_{2}=10\textrm{MHz}$,
$B_{3}=5\textrm{MHz}$; \textbf{(B1>B3>B2):} $B_{1}=15\textrm{MHz}$,
$B_{2}=5\textrm{MHz}$, $B_{3}=10\textrm{MHz}$; \textbf{(B2>B3>B1):} $B_{1}=5\textrm{MHz}$, $B_{2}=15\textrm{MHz}$, $B_{3}=10\textrm{MHz}$; and \textbf{(B3>B2>B1):} $B_{1}=5\textrm{MHz}$, $B_{2}=10\textrm{MHz}$, $B_{3}=15\textrm{MHz}$
\cite{Singh13Offloading}.%
\footnote{There are 6 cases of the allocations of spectra in three-tier heterogeneous networks in total. However, we only analyze 4 typical cases therein. %
} In the following, we will use default values above unless otherwise
declared.

\begin{figure}
  \centering
  \includegraphics[width=6.9cm, height = 6.2cm, draft=false]{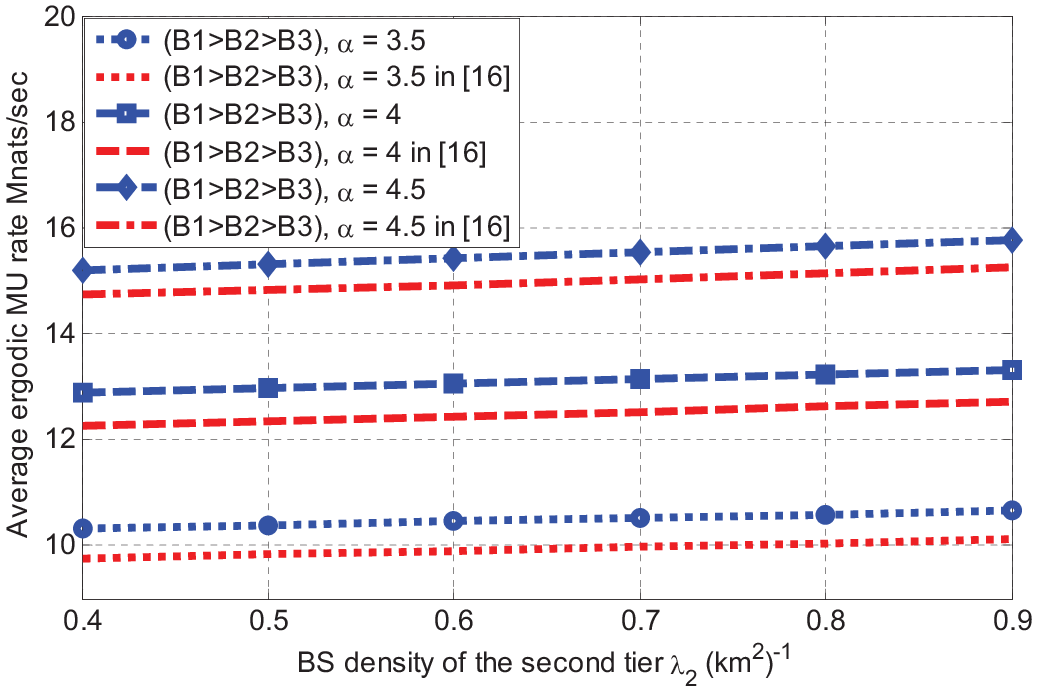}\\
  \caption{\small The average ergodic MU data rate with respect to the density of the 2nd tier BSs with spectral allocation of (B1>B2>B3) in a three-tier heterogeneous network.}\label{fig2}
\end{figure}

Fig.~\ref{fig2} shows the average ergodic MU data rate with respect to the density of the 2nd tier BSs $\lambda_2$ which varies from 0.4$(km^2)^{-1}$ to 0.9$(km^2)^{-1}$ considering three different path loss exponents $\alpha$. We find that the average ergodic MU rate increases slowly with the increasing BS density when we fix the path loss exponent. Also, our proposed cell association scheme outperforms the CRE association scheme analyzed in \cite{Han-Shin12Heterogeneous}, which indicates the effectiveness of ours'. Path loss exponent has more effects on the average ergodic MU data rate when the BS density is fixed. Higher path loss exponent always results in higher average ergodic MU data rate.

\begin{figure}
  \centering
  \includegraphics[width=6.9cm, height = 6.2cm, draft=false]{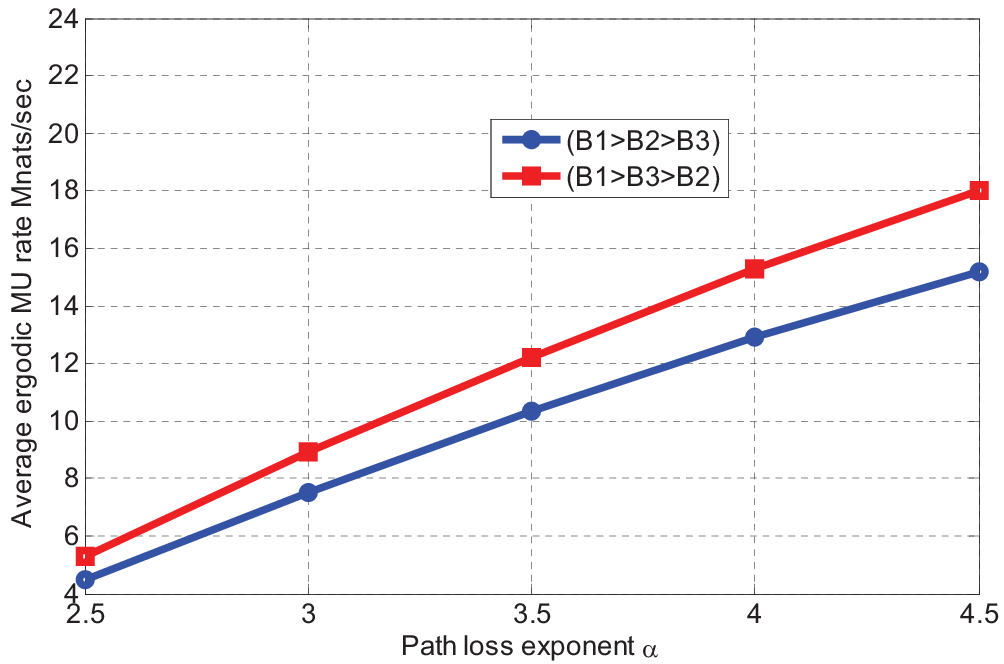}\\
  \caption{\small The average ergodic MU data rate with respect to path loss exponent with two kinds of spectral allocations in a three-tier heterogeneous network.}\label{fig3}
\end{figure}

Fig.~\ref{fig3} compares the average ergodic MU data rate with respect to path loss exponent with two kinds of spectral allocations, i.e., (B1>B2>B3) and (B1>B3>B2). The average ergodic MU data rate increases with the increasing path loss exponent when spectral allocation is fixed, which indicates that to some degree higher path loss exponent contributes network performance. When path loss exponent is fixed, (B1>B3>B2) performs better than (B1>B2>B3). (B1>B2>B3) represents traditional spectral allocation which distributes more spectral resource towards towered BSs, while in (B1>B3>B2) small BSs have more spectral resource than towered BSs. It is implied that if using our proposed cell association scheme, small sized and high-density BSs should be allocated more spectral resource to obtain better holistic performance.

\begin{figure}
  \centering
  \includegraphics[width=6.9cm, height = 6.2cm, draft=false]{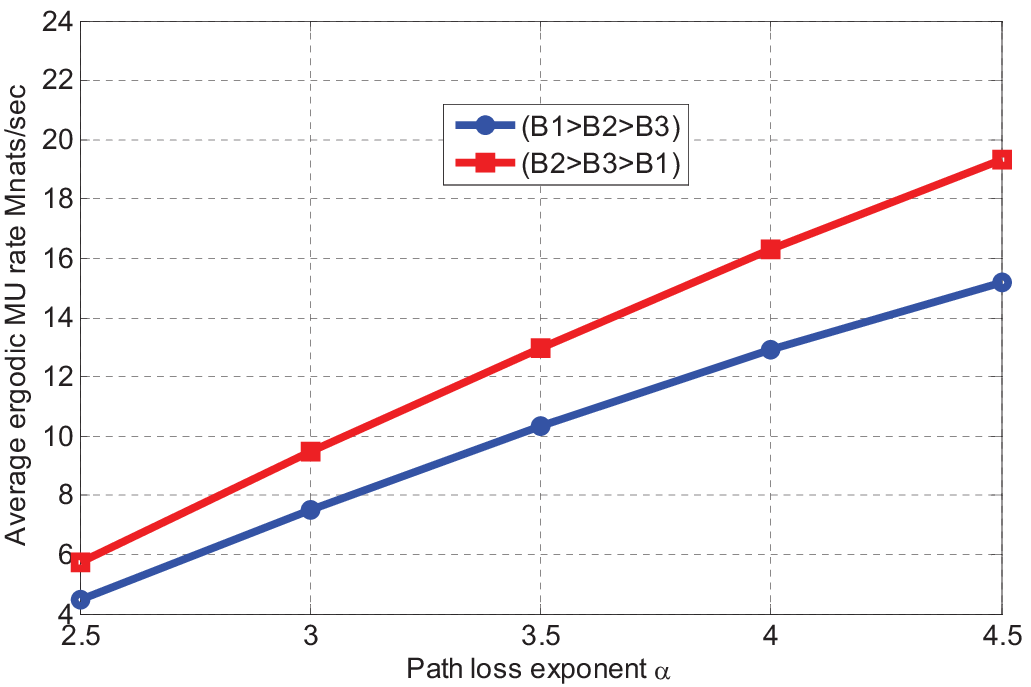}\\
  \caption{\small The average ergodic MU rate with respect to path loss exponent with two kinds of spectral allocations in a three-tier heterogeneous network.}\label{fig4}
\end{figure}

\begin{figure}
  \centering
  \includegraphics[width=6.9cm, height = 6.2cm, draft=false]{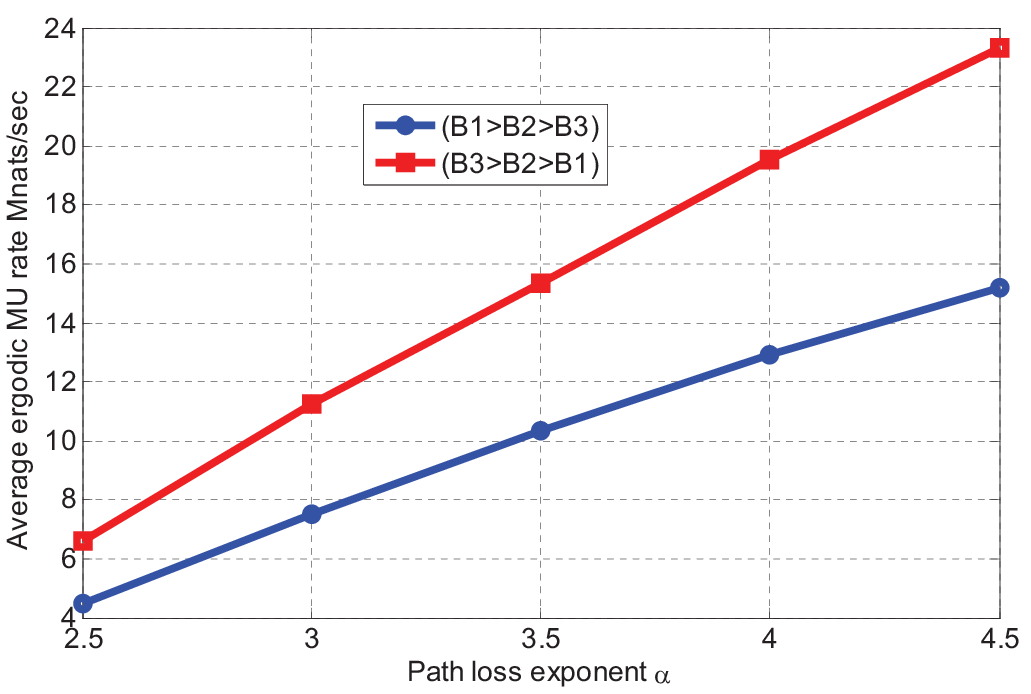}\\
  \caption{\small The average ergodic MU rate with respect to path loss exponent with two kinds of spectral allocations in a three-tier heterogeneous network.}\label{fig5}
\end{figure}

Fig.~\ref{fig4} and fig.~\ref{fig5} illustrate the average ergodic MU data rate with respect to path loss exponent with two kinds of spectral allocations. We obtain similar conclusions obtained from fig.~\ref{fig3}. However, the gap between the two curves of (B1>B2>B3) and (B2>B3>B1) in fig.~\ref{fig4} and the gap between the two curves of (B1>B2>B3) and (B3>B2>B1) in fig.~\ref{fig5} are bigger than that in fig.~\ref{fig3}, which again indicates that small sized and high-density BSs should be allocated more spectral resource if using our proposed cell association scheme.

\section{Conclusions\label{sec:Conclusion-and-future}}

In this paper, motivated by the problems of existing cell association schemes which are merely based on one indicator like RSS, SINR or distance from nearby BSs, we propose a new cell association scheme by joint consideration of SINR and traffic load in heterogeneous networks. Through numerical results, we find that our proposed cell association scheme outperforms CRE association scheme. Also, the results provide some insights of spectral allocation by using our proposed cell association scheme, which implies that allocating small sized and high-density BSs more spectral resource results in better holistic performance.

Still, some work need to be done to further this proposed cell association scheme. For instance, if the number of candidate BSs $n$ in each tier network is more than one, the corresponding analysis may be more general. And also, adding shadowing may make the scenario more realistic.

\section*{Acknowledgment}

The authors would like to acknowledge the support from the International Science and Technology Cooperation Program of China (Grant No. 2014DFA11640 and 2012DFG12250), the National Natural Science Foundation of China (NSFC) (Grant No. 61271224, 61471180 and 61301128), NFSC Major International Joint Research Project (Grant No. 61210002), the Hubei Provincial Science and Technology Department (Grant No. 2013BHE005), the Fundamental Research Funds for the Central Universities (Grant 2013QN136, 2014QN155 and 2013ZZGH009), and EU FP7-PEOPLE-IRSES (Contract/Grant No. 247083, 318992 and 610524).

%\section*{Acknowledgment}
%
%The authors would like to acknowledge the support from the International Science and Technology Cooperation Program of China under the grants 2014DFA11640 and 0903, National Natural Science Foundation of China (NSFC) under the grants 61271224 and 61301128, NFSC Major International Joint Research Project under the grant 61210002, the Hubei Provincial Science and Technology Department under the grant 2013BHE005, the Fundamental Research Funds for the Central Universities under the grant 2011QN020 and 2013QN136, and Special Research Fund for the Doctoral Program of Higher Education (SRFDP) under the grant 20130142120044. This research is partially supported by EU FP7-PEOPLE-IRSES, project acronym S2EuNet (grant no. 247083), project acronym WiNDOW (grant no. 318992) and project acronym CROWN (grant no. 610524). Guoqiang Mao's work is supported by Australian Research Council Discovery projects DP110100538 and DP120102030. And the authors also appreciate the help of numerical results by Junliang Ye, Jing Yang and Guoqi Zou.

\bibliographystyle{IEEEtran}
\addcontentsline{toc}{section}{\refname}\bibliography{ICC2015}

\end{document}